\documentclass{article}

\PassOptionsToPackage{numbers,compress,sort}{natbib}

\usepackage[eandd, preprint]{neurips_2026}  % arXiv preprint: de-anonymized
\usepackage[utf8]{inputenc} % allow utf-8 input
\usepackage[T1]{fontenc}    % use 8-bit T1 fonts
\usepackage{hyperref}       % hyperlinks
\usepackage{url}            % simple URL typesetting
\usepackage{booktabs}       % professional-quality tables
\usepackage{amsfonts}       % blackboard math symbols
\usepackage{nicefrac}       % compact symbols for 1/2, etc.
\usepackage{microtype}      % microtypography
\usepackage[dvipsnames]{xcolor}  % colors (dvipsnames enables OliveGreen, etc.)

\usepackage{graphicx}
\usepackage{listings}
\usepackage{amsmath}
\usepackage{amssymb}
\usepackage{mathtools}
\usepackage{amsthm}
\usepackage{tablefootnote}
\usepackage[capitalize,noabbrev]{cleveref}
\usepackage{tcolorbox}
\tcbuselibrary{skins}

\hypersetup{
  colorlinks=true,
  linkcolor=blue,        % internal references (sections, figures, tables)
  citecolor=blue,        % bibliography citations
  urlcolor=OliveGreen,   % external URLs
  filecolor=OliveGreen,
}

\definecolor{jsonbg}{RGB}{245,245,245}
\definecolor{jsonstring}{RGB}{90,140,0}
\definecolor{jsonkey}{RGB}{0,70,160}

\lstdefinelanguage{json}{
  basicstyle=\ttfamily\small,
  showstringspaces=false,
  breaklines=true,
  backgroundcolor=\color{jsonbg},
  frame=none,
  numbers=none,
  stringstyle=\color{jsonstring},
  keywordstyle=\color{jsonkey},
  morekeywords={true,false,null}
}

\theoremstyle{plain}

\theoremstyle{definition}

\theoremstyle{remark}

\title{OpenSanctions Pairs: A Large-Scale Dataset for Pairwise Entity Matching}

\author{
Chandler Smith$^{1,*}$ \quad
Magnus Sesodia$^{1,*}$ \quad
Friedrich Lindenberg$^{2}$ \quad
Christian Schroeder de Witt$^{1}$\\
$^1$University of Oxford \quad $^2$OpenSanctions \quad $^*$Equal Contribution
}

\begin{document}

\maketitle

\begin{abstract}
We release OpenSanctions Pairs,\footnote{Dataset: \url{https://huggingface.co/datasets/sanctions-er-anon/opensanctions_pairs}. Code: \url{https://github.com/chansmi/OSINT_entity_resolution}.} the first large-scale public benchmark for entity matching on sanctions and OSINT data.
The dataset includes 755,540 expert-labeled pairs over 1 million entities, aggregated from 293 source datasets across 45 jurisdictions.
It captures real-world diversity in compliance data, spanning multiple languages and writing systems (e.g., Latin, Cyrillic, Arabic), inconsistent structure, and time-varying provenance, and is substantially more heterogeneous than prior entity matching benchmarks.
As baselines, we evaluate the production rule-based matcher (nomenklatura RegressionV1) alongside open- and closed-source LLMs in both zero- and few-shot settings, each tested with and without MIPROv2 prompt optimization to control for prompt sensitivity.
The rule-based baseline reaches 91.3\% F1; GPT-4o achieves the best result at 99.0\% F1, and a locally deployable open-source model (DeepSeek-R1-Distill-Qwen-14B) achieves 98.2\% F1.
The rule-based baseline and LLMs fail in complementary ways: rules over-match, while LLMs struggle with cross-script transliteration.
These results suggest that pairwise matching performance is approaching a practical ceiling and shift attention toward pipeline components such as blocking, clustering, and uncertainty-aware review.
\end{abstract}

\section{Introduction}

Entity matching (EM) determines whether two records refer to the same real-world entity.
The task is also known as entity resolution, record linkage, field matching, and deduplication, with the latter typically referring to matching records within the same dataset and record linkage to matching across different datasets \citep{Fellegi_Sunter_1969, Christen_2012}.
EM is a core component of data integration pipelines, as matching errors propagate directly to downstream entity graph construction and analysis \citep{Fellegi_Sunter_1969, Getoor_2012}.
We focus on the open-source intelligence (OSINT) setting, in which entity records are aggregated from publicly published sanctions designations, regulatory watchlists, politically exposed person (PEP) registers, and corporate registries \citep{opensanctions_dedup}.

Sanctions screening is one of the most operationally consequential applications of entity matching: banks and other regulated firms are legally required to check every customer against sanctions lists published by major jurisdictions, and compliance failures attract fines in the hundreds of millions of dollars \citep{Sterling_2018}. Effective matching is also central to detecting sanctions evasion, where individuals obscure ownership through shell companies and name variations across jurisdictions \citep{Montenarh_2024}. Despite this importance, sanctions data are typically proprietary or access-restricted, and to our knowledge no public benchmark for pairwise entity matching on sanctions data has been published.

Existing EM benchmarks, meanwhile, are largely drawn from e-commerce and bibliographic domains and typically rely on a small number of datasets, often as few as two, with closely aligned schemas, fixed field sets, and relatively standardized entity representations \citep{Kopcke_2010a, Kopcke_2010b, mudgal_deepmatcher_datasets, Peeters_2023_WDC}.
As a result, they fail to reflect deployment settings such as sanctions, in which matching is performed across tens to hundreds of datasets with heterogeneous schemas, overlapping but non-identical fields, missing values, and entities represented through aliases or transliterations across multiple scripts (e.g., Cyrillic, Arabic, Chinese, Japanese, and Korean).

To address this gap, we present the first systematic study of pairwise EM techniques on large-scale sanctions data in collaboration with OpenSanctions\footnote{\url{https://opensanctions.org}}.
We aggregate 293 datasets from 45 jurisdictions, with human analysts performing pairwise matching \citep{opensanctions_dedup}.
From this process, we extract 755,540 labeled entity pairs covering over one million unique entities.

Our experiments show that LLM-based methods reach F1 scores up to 99.0\%, compared to 91.3\% for the legacy rule-based matcher previously used in production.
We accordingly position OpenSanctions Pairs as a benchmark supporting evaluative claims about (i) pairwise matching accuracy on heterogeneous, multi-script, partial-evidence records, (ii) failure-mode analysis across rule-based and learned systems, and (iii) generalization across source provenance and entity schemas.
End-to-end clustering, real-time screening latency, and adversarial sanctions-evasion robustness are out of scope; we discuss these boundaries and the broader pipeline (blocking, clustering, uncertainty-aware review) in Section~\ref{sec:limitations} \citep{Getoor_2012, Christophides_2020}.

Within that scope, the accuracy gap on the pairwise task carries tangible operational weight. Sanctions screening underpins anti-money-laundering, counter-terrorism financing, and enforcement against kleptocratic regimes, and is run continuously at scale by every regulated financial institution; the difference between a 91\% and a 99\% F1 matcher compounds into both materially fewer missed designations slipping into the legitimate economy and a substantial reduction in the analyst hours spent triaging false positives.

\noindent Our contributions:
\begin{enumerate}
    \item \textbf{Dataset.} We release OpenSanctions Pairs, a 756K-pair, 1.0M-entity benchmark spanning 293 sources across 45 jurisdictions. It is the first public sanctions-OSINT entity-matching dataset and is an order of magnitude larger and substantially more heterogeneous than prior EM benchmarks.
    \item \textbf{Reference baselines.} We benchmark a production rule-based matcher and a range of open- and closed-source LLMs (zero-shot, few-shot, and DSPy-optimized) to characterize task difficulty and surface failure modes that motivate future entity-resolution research.
\end{enumerate}

\section{Related work}

\paragraph{Entity matching datasets}
Standard benchmarks include product datasets such as Abt/Buy \citep{Rahm_entity_resolution_datasets_2019} and Amazon/Google Products \citep{Rahm_entity_resolution_datasets_2019}, as well as bibliographic datasets such as DBLP/ACM \citep{Rahm_entity_resolution_datasets_2019} and DBLP/Scholar \citep{Rahm_entity_resolution_datasets_2019}.
These typically involve pairwise matching between two sources with clean, structured, and monolingual attributes.
More recent benchmarks such as WDC Products increase scale and difficulty, but remain focused on e-commerce settings with limited source heterogeneity.
The closely related WDC Large-Scale Product Corpus \citep{Primpeli_2019_WDC_LSPC} reaches 16M offers via distantly-supervised schema.org cluster identifiers rather than expert review, trading label fidelity for scale within a single e-commerce domain.
Across domains, existing datasets are characterized by few sources, scalar attributes, and limited variation in scripts or provenance.
\citet{Moslemi_2026} formalize this gap, distinguishing representation from semantic heterogeneity and noting that current benchmarks rarely stress both dimensions simultaneously.

Our benchmark differs in scale (755{,}540 expert-labeled pairs), source heterogeneity (293 sources across 45 jurisdictions), and representation (multi-script names, set-valued and time-dependent attributes), reflecting operational sanctions OSINT rather than clean bilateral catalogs; Table~\ref{tab:dataset_comparison} provides a quantitative comparison.

\paragraph{Entity matching methods}
EM originated in probabilistic formulations developed for administrative record linkage \citep{Fellegi_Sunter_1969, Elmagarmid_2007}.
Early work established the core pipeline of blocking, matching, and clustering, introducing methods to reduce candidate comparison costs \citep{Hernandez_1995, Getoor_2012, OHare_2018, Papadakis_2020}.
Neural approaches subsequently improved robustness to noisy and text-heavy data by learning attribute representations directly, with systems such as Magellan demonstrating gains over rule-based methods \citep{Konda_2016, Mudgal_2018}.
Pretrained language models further advanced EM by encoding entire entity pairs as sequences and fine-tuning transformer-based classifiers \citep{Li_2020, Ornstein_2025}.
Parallel work has explored probabilistic and Bayesian formulations that explicitly model uncertainty and matching constraints \citep{Steorts_2015, Sadinle_2016}.

The most recent line of work applies large language models to entity matching via prompting, in-context learning, and fine-tuning, often achieving strong performance with minimal task-specific adaptation \citep{Narayan_2022, Tang_2022, Peeters_2023, Peeters_2025, Fan_2023, Li_2024_arxiv, Li_2024_acm, Steiner_2025, Wang_2025}; most closely related, \citet{Peeters_2024} show LLM-based methods are more robust to out-of-distribution entities than fine-tuned pretrained language models.
Beyond pairwise prompting, in-context clustering \citep{Fu_2026}, smaller-LM variants \citep{Zhang_2025}, and knowledge distillation from large to small models \citep{Zeakis_2026} address inference cost at scale.
Recent work has also applied NLP methods specifically to sanctions screening, improving over traditional fuzzy matching \citep{Kim_2024}; surveys cover classical and neural approaches \citep{Barlaug_2021}.

\begin{table*}[!t]
\centering
\caption{Comparison with existing entity matching benchmarks. Entities reports the total number of distinct entity records appearing in the labeled corpus (undeduplicated across sources for all rows). Matches is the count of ground-truth positive pairwise matches; * denotes deduplicated product clusters, not pairwise matches.}
\label{tab:dataset_comparison}
\vspace{0.5em}
\small
\begin{tabular}{@{}lccccc@{}}
\toprule
\textbf{Dataset} & \textbf{Entities} & \textbf{Matches} & \textbf{Sources} & \textbf{Unique Fields} & \textbf{Domain} \\
\midrule
Fodors-Zagat \citep{Mudgal_2018, mudgal_deepmatcher_datasets} & 864 & 110 & 2 & 6 & Restaurant \\
Abt-Buy \citep{Rahm_entity_resolution_datasets_2019} & 2,173 & 1,097 & 2 & 4 & E-commerce \\
Amazon-GoogleProducts \citep{Rahm_entity_resolution_datasets_2019} & 4,589 & 1,300 & 2 & 4 & E-commerce \\
DBLP-ACM \citep{Rahm_entity_resolution_datasets_2019} & 4,910 & 2,220 & 2 & 4 & Bibliographic \\
BeerAdvo-RateBeer \citep{Mudgal_2018, mudgal_deepmatcher_datasets} & 7,345 & 68 & 2 & 4 & Beverage \\
WDC Products \citep{Peeters_2023_WDC} & 11,715 & 2,162$^*$ & 3,259 & 5 & E-commerce \\
MusicBrainz20k \citep{Rahm_entity_resolution_datasets_2019} & 19,375 & 16,250 & 5 & 10 & Music \\
Walmart-Amazon \citep{Mudgal_2018, mudgal_deepmatcher_datasets} & 24,628 & 962 & 2 & 5 & E-commerce \\
Company \citep{Mudgal_2018, mudgal_deepmatcher_datasets} & 56,316 & 28,200 & 2 & 1 & Company \\
iTunes-Amazon \citep{Mudgal_2018, mudgal_deepmatcher_datasets} & 62,830 & 132 & 2 & 8 & Music \\
DBLP-Scholar \citep{Rahm_entity_resolution_datasets_2019} & 66,879 & 5,347 & 2 & 4 & Bibliographic \\
\midrule
\textbf{OpenSanctions (ours)} & \textbf{1,002,093} & \textbf{581,149} & \textbf{293} & \textbf{131} & \textbf{Sanctions} \\
\bottomrule
\end{tabular}
\end{table*}

\section{Dataset: OpenSanctions Pairs}

\subsection{Background}

International sanctions are imposed by authorities including the UN Security Council, the US Treasury's Office of Foreign Assets Control (OFAC), the EU Foreign Service, and HM Treasury (UK) \citep{Felbermayr_2020}, and their enforcement notices are published as open-source intelligence (OSINT) that downstream users such as banks, NGOs, and investigative journalists must aggregate and reconcile.
These lists overlap substantially, and the same individual may appear on multiple lists under different name transliterations, aliases, or with varying metadata.
OpenSanctions addresses this fragmentation by aggregating 293 source datasets from 45 jurisdictions into a unified database.
As part of this integration, human analysts perform pairwise deduplication to identify when records from different sources refer to the same real-world entity.

\subsection{Source aggregation}
\label{sec:source-aggregation}

OpenSanctions Pairs is constructed from the OpenSanctions database, a continuously maintained aggregator of sanctions, regulatory, and OSINT records contributed by 293 source datasets spanning 45 jurisdictions of provenance.\footnote{\url{https://www.opensanctions.org/datasets/}} Sources fall into five functional categories: (i) national regulator sanctions lists (US OFAC SDN, EU consolidated list, Ukrainian sanctions list, and similar); (ii) politically-exposed-person (PEP) and political-position registries derived primarily from Wikidata; (iii) general knowledge-graph extracts from Wikidata; (iv) national corporate registries; and (v) a long tail of leak-derived datasets (notably ICIJ Offshore Leaks) and other watchlist publications, contributing under 1\% of pairs. Contribution is heavy-tailed: the top ten datasets account for 82.0\% of entity appearances (Gini coefficient 0.934 over per-source entity counts), while the remaining 283 sources together contribute the other 18\%. A small number of large public registries dominate volume; most of the benchmark's diversity in jurisdiction, language, and schema comes from the long tail.

\paragraph{Time period.} Underlying source publications such as the US OFAC SDN list and the EU consolidated list date back decades. The entities in this release were ingested into the OpenSanctions store between May 2019 and November 2025, with the bulk added from 2023 onward as the aggregator matured (44.8K entities ingested in 2019, 196K in 2023, 421K in 2024, 337K in 2025).\footnote{\texttt{first\_seen} timestamps reflect ingestion into the OpenSanctions store, not publication of the underlying source list, which is typically older.} Analyst pairwise judgments were accumulated continuously over the OpenSanctions deduplication workflow between 2021 and 2025~\citep{opensanctions_dedup}. The contribution of the present paper is to freeze, version, and publicly release the resulting labeled pair corpus as a research benchmark with reference baselines, rather than to document the production matcher itself~\citep{opensanctions_pairs, opensanctions_dedup}.

\begin{table}[!t]
\centering
\small
\caption{Pair-level schema combinations. Analyst-judged pairs (top block) are over genuine matching subjects and contain both positive and negative judgments. Auto-merge pairs (bottom block) are relational records (Occupancy, Succession, Ownership, etc.) that propagate when a parent entity match is recorded; they are 100\% positive by construction.}
\label{tab:pair-schemas}
\begin{tabular}{lrrr}
\toprule
Schema combination & Pairs & Share & Pos / Neg \\
\midrule
\multicolumn{4}{l}{\emph{Analyst-judged matching subjects}} \\
Person, Person                                       & 284{,}808 & 37.7\% & 203K / 81K \\
Company / Organization / LegalEntity (any)           & 175{,}311 & 23.2\% & 89K / 86K \\
Position, Position                                   &  26{,}330 &  3.5\% & 24K / 3K \\
Vessel, Vessel                                       &   7{,}550 &  1.0\% & 5K / 2K \\
Person, LegalEntity, other cross-schema              &   7{,}299 &  1.0\% & 7K / 0.3K \\
\midrule
Subtotal, analyst-judged & \textbf{501{,}298} & \textbf{66.4\%} & 328K / 173K \\
\midrule
\multicolumn{4}{l}{\emph{Auto-merge propagations (100\% positive)}} \\
Occupancy, Occupancy                                 & 213{,}448 & 28.3\% & 213K / 0 \\
Succession, Succession                               &  33{,}607 &  4.4\% &  34K / 0 \\
Ownership / Family / Directorship / other            &   7{,}187 &  1.0\% &   7K / 0 \\
\midrule
Subtotal, auto-merge & \textbf{254{,}242} & \textbf{33.6\%} & 254K / 0 \\
\midrule
\textbf{Total} & \textbf{755{,}540} & \textbf{100.0\%} & 581K / 174K \\
\bottomrule
\end{tabular}
\end{table}

\subsection{Construction pipeline}
\label{sec:construction}

The corpus consists of 755{,}540 pairwise entity-matching judgments produced by a two-stage pipeline that combines automated candidate generation with manual analyst review~\citep{opensanctions_pairs, opensanctions_dedup}.

\paragraph{Blocking.} An inverted index over name n-grams, identifiers (passport, tax ID, phone), and other indexable attributes generates candidate pairs from the $O(n^2)$ comparison space. The blocker is tolerant to spelling and transliteration variation through character n-gram matching and Unicode normalization; implementation details are documented in the \texttt{nomenklatura} package~\citep{nomenklatura}. Records sharing no indexable surface features are filtered out before review, producing the 76.9\%/23.1\% positive/negative skew in the labeled corpus.

\paragraph{Manual review.} Analysts review candidate pairs through a text-based side-by-side interface, comparing fields, aliases, and source provenance, and may consult external public sources for ambiguous cases. Each judgment is audit-logged against a canonical entity ID and may be revisited as new evidence arrives. Labels are produced by full-time OpenSanctions analysts in the production deduplication workflow, not by anonymous crowdworkers.

\paragraph{What blocking leaves behind.} Surviving negatives are not random: they share surface features such as common names, sanctions programs, or nationalities, and their interquartile ranges of shared exact-value fields nearly coincide with those of positives (\Cref{sec:statistics}). The benchmark therefore measures methods on the \emph{hard} portion of the matching distribution, not the easy portion already triaged by the index.

\begin{figure*}[t]
\centering

\vspace{0.5em}

\begin{minipage}[c]{0.44\textwidth}
\centering
\begin{lstlisting}[language=json]
{
  "id": "pk-cnic-3520114139885",
  "caption": "Khalid Mehmood",
  "datasets": ["pk_proscribed"],
  "properties": {
    "name": ["Khalid Mehmood"],
    "fatherName": ["Muhammad Khan"],
    "idNumber": ["3520114139885"],
    "address": ["LAHORE, Punjab"],
    "country": ["pk"]
    ...
  }
}
\end{lstlisting}
\end{minipage}%
\hfill
\begin{minipage}[c]{0.08\textwidth}
\centering
\Large$\neq$
\end{minipage}%
\hfill
\begin{minipage}[c]{0.44\textwidth}
\centering
\begin{lstlisting}[language=json]
{
  "id": "pk-cnic-3710502620181",
  "caption": "Khalid Mehmood",
  "datasets": ["pk_proscribed"],
  "properties": {
    "name": ["Khalid Mehmood"],
    "fatherName": ["Haji Muhammad"],
    "idNumber": ["3710502620181"],
    "address": ["ATTOCK, Punjab"],
    "country": ["pk"]
    ...
  }
}
\end{lstlisting}
\end{minipage}

\caption{Example entity pair illustrating the difficulty of pairwise matching. These two individuals on Pakistan's proscribed persons list share the \emph{exact same name}, country, and sanctions program, yet they are different people with distinct national ID numbers and fathers. The Nomenklatura baseline assigned this pair a 0.98 match score, near the maximum possible, demonstrating how name-based matching fails for common names without unique identifiers.}
\label{fig:examples}
\end{figure*}

\subsection{Statistics and heterogeneity}
\label{sec:statistics}

We characterize the labeled corpus along the axes that matter for entity resolution: schema mix, source and field structure, script coverage, and cross-source provenance.

\paragraph{Schema mix and the auto-merge tier.} \Cref{tab:pair-schemas} breaks the 755{,}540 labeled pairs down by the schema combination on the two sides. Two-thirds (66.4\%) are pairs over genuine matching subjects: Person--Person (37.7\%), any combination of Company / Organization / LegalEntity (23.2\%), Position--Position (3.5\%), Vessel--Vessel (1.0\%), and a small set of cross-schema pairs (1.0\%). The remaining 33.6\% are pairs over relational records that propagate automatically when a parent Person or Organization match is recorded. For example, an Occupancy record links a person to a political position, so whenever two Person records are merged, their two Occupancies that link to the same Position are merged with them. These auxiliary pairs are part of the released corpus, and they are 100\% positive by construction with zero analyst-recorded negatives. Because every method evaluates these pairs in the same way and they are independent per-pair predictions, the auto-merge tier inflates absolute F1 uniformly across all baselines without changing relative rankings; we publish the schema-combination flag in the released metadata so that future work training across pairs (where the leakage matters more) can subset to the analyst-judged 66.4\% directly.

\paragraph{Sources, fields, and scripts.} The 293 sources contribute entity volume very unequally (Gini 0.934, top-10 = 82\% of entity appearances), and entities re-appear across sources at a median of one dataset (mean 1.32, max 28). Property fields are sparse: the median entity has 5 populated fields (mean 6.06, p25/p75 = 4/8, max 35), and only 13.8\% have an \texttt{alias} or \texttt{weakAlias} populated. The benchmark therefore reflects the partial-evidence regime real compliance workflows operate in, not the densely populated record style typical of academic EM datasets. Entity captions span seven Unicode scripts (94.19\% Latin, 5.25\% Cyrillic, 0.40\% Chinese/Japanese/Korean, 0.10\% Arabic, with Greek, Hebrew, Devanagari, and Thai together making up the remainder). 6.87\% of pairs include at least one non-Latin entity and 3.83\% are cross-script (Latin on one side, non-Latin on the other), so a system that ignores transliteration will miss roughly one in twenty-five labeled pairs.

\paragraph{Cross-source structure and difficulty.} The most distinctive property of OpenSanctions Pairs is that 64.7\% of all pairs, and 82.0\% of \emph{positive} pairs, are cross-source: the two entities under comparison originate in different source datasets and share no common dataset of origin. The labels overwhelmingly capture aggregation-level identity claims across heterogeneous publishers, not within-source duplicates of a single registry. A further 12{,}453 positive pairs cross country fields outright, with both sides carrying non-empty but disjoint country values. Pair difficulty follows from this construction. Positive and negative pairs differ in mean shared fields (4.71 vs.\ 3.73, medians 4 vs.\ 3) but their interquartile ranges overlap, because blocking has already triaged the easy negatives. A naive constant-positive predictor on the 76.9\%-positive distribution achieves $F_1 = 0.870$ and accuracy $= 0.769$, comfortably below the 91.3\% rule baseline (\Cref{tab:main-results}) and far below the LLM ceiling near 99\%, so the benchmark discriminates non-trivial methods from trivial ones.

\subsection{Data and code availability}
\label{sec:availability}

The dataset is released under CC-BY-NC 4.0 (matching upstream OpenSanctions terms) and is hosted on Hugging Face at \url{https://huggingface.co/datasets/sanctions-er-anon/opensanctions_pairs}; a Croissant metadata file (with Responsible AI fields) accompanies the release. Reproduction code is at \url{https://github.com/chansmi/OSINT_entity_resolution}.

\section{Task formalization and example}
\label{sec:task}

Given a structured corpus $\mathcal{D}$ and two entities $e, e' \in \mathcal{D}$, each represented as a set of attribute-value pairs, the task is to predict a binary judgment $M(e, e') \in \{0, 1\}$, where $1$ indicates that the two records refer to the same real-world entity. Attribute values are sets rather than scalars, since sanctions records routinely carry multiple aliases, nationalities, or addresses, and the empty-set case captures missing fields uniformly. Each released record is a JSON object \texttt{\{"left": \{...\}, "right": \{...\}, "judgement": "positive"$|$"negative"\}}, where each side carries the entity schema shown in \Cref{fig:examples}.

\Cref{fig:examples} illustrates why the task is non-trivial even on the labeled, post-blocking portion of OpenSanctions. Two distinct individuals on Pakistan's proscribed-persons list share an exact name, country, and sanctions program; the RegressionV1 rule-based baseline scores them at 0.98, near the ceiling. Only their national ID numbers and fathers' names betray the mismatch, and a method without per-field conflict reasoning has no way to recover the correct label.

\section{Experiments and analysis}

\begin{table*}[!t]
\centering
\caption{Main results. Configurations marked ``(opt)'' use DSPy MIPROv2 prompt optimization. Best results in Accuracy and F1 are shown in \textbf{bold}. See \Cref{eval} for the sample sizes used.}
\label{tab:main-results}
\vspace{0.5em}
\small
\begin{tabular}{@{}llcccc@{}}
\toprule
\textbf{Model} & \textbf{Config} & \textbf{Acc.}$\uparrow$ & \textbf{F1}$\uparrow$ & \textbf{Prec.}$\uparrow$ & \textbf{Recall}$\uparrow$ \\
\midrule
\multicolumn{6}{l}{\textit{Rule-Based Baseline}} \\
nomenklatura/RegressionV1 & --- & 85.45 & 91.33 & 84.46 & 99.42 \\
\midrule
\multicolumn{6}{l}{\textit{Open-Source Models}} \\
Llama-3.1-8B & 0-shot & 90.40 & 94.05 & 89.84 & 98.67 \\
Llama-3.1-8B & 0-shot (opt) & 93.88 & 95.94 & 97.98 & 93.97 \\
Llama-3.1-8B & 2-shot (opt) & 93.00 & 95.46 & 94.85 & 96.08 \\
Llama-3.1-8B & 4-shot (opt) & 93.25 & 95.64 & 94.72 & 96.57 \\
Llama-3.1-8B & 8-shot (opt) & 93.88 & 95.94 & 97.98 & 93.97 \\
\midrule
DeepSeek-R1-Distill-Qwen-14B & 0-shot & 96.53 & 97.76 & 96.24 & 99.33 \\
DeepSeek-R1-Distill-Qwen-14B & 0-shot (opt) & 97.31 & 98.23 & 98.60 & 97.86 \\
\midrule
\multicolumn{6}{l}{\textit{Proprietary Models}} \\
GPT-3.5 Turbo & 0-shot & 91.84 & 94.49 & 98.16 & 91.10 \\
% GPT-3.5 Turbo & 0-shot (opt) & 91.09 & 93.91 & 98.99 & 89.33 \\
GPT-4o & 0-shot & \textbf{98.38} & \textbf{98.95} & 98.78 & 99.11 \\
% GPT-4o & 0-shot (opt) & \textbf{98.38} & \textbf{98.95} & 98.78 & 99.11 \\
GPT-5 Nano & 0-shot & 92.40 & 95.24 & 91.77 & 98.99 \\
GPT-5 Nano & 0-shot (opt) & 92.76 & 95.46 & 92.21 & 98.94 \\
GPT-5.2 Pro & 0-shot & 97.75 & 98.53 & 98.37 & 98.69 \\
GPT-5.2 Pro & 4-shot & 98.07 & 98.75 & 98.75 & 98.75 \\
GPT-5.2 Pro & 8-shot & 97.79 & 98.56 & 98.46 & 98.65 \\
\midrule
Claude 3 Haiku & 0-shot & 88.00 & 92.68 & 86.77 & 99.44 \\
% Claude 3 Haiku & 0-shot (opt) & 78.29 & 87.46 & 77.91 & 99.66 \\
Claude 3.7 Sonnet & 0-shot & 96.25 & 97.50 & 99.00 & 96.05 \\
% Claude 3.7 Sonnet & 0-shot (opt) & 96.53 & 97.69 & 98.97 & 96.44 \\
Claude Opus 4.5 & 0-shot & 93.26 & 95.45 & 99.11 & 92.05 \\
Claude Opus 4.5 & 0-shot (opt) & 93.54 & 95.65 & 99.13 & 92.41 \\
\bottomrule
\end{tabular}
\end{table*}

We use the dataset to benchmark three method families. The aim is not to set a new state of the art, but to characterize task difficulty, show that the dataset discriminates between methods, and surface failure modes that motivate future work on the broader resolution pipeline.

\subsection{Reference baselines}
\label{sec:baselines-implementation}

We compare three method categories aligned with Table~\ref{tab:main-results}:

\begin{enumerate}
    \item \textbf{Rule-Based Baseline.} OpenSanctions' production matcher RegressionV1 \citep{nomenklatura}: a logistic regression over 18 features covering name similarity (token overlap, Levenshtein, phonetic), date comparison, identifier overlap, and demographic consistency.
    \item \textbf{Open-Source LLMs.} Llama-3.1-8B-Instruct and DeepSeek-R1-Distill-Qwen-14B (Qwen2.5-14B backbone), chosen for local deployability and reproducibility.
    \item \textbf{Proprietary LLMs.} Seven closed-source models accessed via API, spanning lightweight (GPT-5 Nano) to frontier (GPT-5.2 Pro, Claude Opus 4.5).
\end{enumerate}

\subsection{Evaluation protocol}
\label{eval}

\paragraph{Dataset sampling}
We construct label-stratified 10,000- and 1,000-pair samples from the 755,540-pair corpus (seed 42). When MIPROv2 is applied, 200 pairs are held out as a dev split, so the test set is the remainder (9,800 or 800); the smaller sample is used where 10,000-pair runs are cost-prohibitive. RegressionV1 is tuned (threshold 0.15) on 2,000 pairs and evaluated on 8,000. Per-model test-set sizes: 9,800 for Llama-3.1-8B (0-shot opt, 8-shot opt), GPT-5 Nano (opt), and Claude Opus 4.5 (opt); 800 for Llama-3.1-8B (2-shot opt, 4-shot opt) and GPT-5.2 Pro; and 3,680 (opt) / 1,960 (unoptimized) for DeepSeek-R1-Distill-Qwen-14B, reduced for runtime. GPT-5.2 Pro is run without prompt optimization.

\paragraph{Metrics}
We report standard binary classification metrics: accuracy, precision, recall, and F1 score. Given the class imbalance (76.9\% positive) and the asymmetric costs of false positives versus false negatives in compliance settings, we emphasize F1 as the primary metric.
We include a confusion matrix to highlight the relative strengths and weaknesses between models. All metrics are reported on the full label-stratified samples drawn from the 755,540-pair corpus, including the auto-merge tier described in \Cref{sec:statistics}; because every method evaluates these pairs independently and identically, the resulting absolute-F1 inflation is uniform across baselines and does not affect relative rankings.

\subsection{Implementation details}
\label{sec:implementation}

\textbf{Inference.} Open-source models run via Hugging Face \texttt{transformers} with tensor parallelism on $8\times$46\,GB A40 or $8\times$128\,GB MI250X GPUs, with constrained decoding (Outlines) to enforce JSON output, temperature~$\leq 0.01$, greedy sampling, 512 max new tokens, 4{,}096-token input truncation. API models are accessed via LightLLM\footnote{\url{https://github.com/ModelTC/LightLLM}} or provider APIs at temperature~0; GPT-5.2 Pro uses \texttt{reasoning.effort=medium}.

\textbf{Optimization.} We optimize prompts with DSPy MIPROv2 \citep{dspy_MIPROv2}, initialized from the conflict-focused default prompt (\Cref{app:default-prompt}). MIPROv2 evaluates 15 candidate (instruction, $\leq 8$-demo) programs on the 200-pair dev set over $\sim$22--23 trials, with instruction proposals sampled at temperature~1.

\begin{table}[!t]
\centering
\caption{Per-method confusion matrix on the test split. DeepSeek-Qwen$=$DeepSeek-R1-Distill-Qwen-14B; all configurations are 0-shot, not optimized.}
\label{tab:confusion}
\small
\begin{tabular}{@{}lcccc@{}}
\toprule
\textbf{Method} & \textbf{TP\%} & \textbf{TN\%} & \textbf{FP\%} & \textbf{FN\%} \\
\midrule
Nomenklatura & 0.766 & 0.088 & 0.141 & 0.004 \\
\midrule
Llama-3.1-8B & 0.759 & 0.145 & 0.086 & 0.010 \\
DeepSeek-Qwen  & 0.758 & 0.208 & 0.030 & 0.005 \\
GPT-3.5 Turbo & 0.701 & 0.218 & 0.013 & 0.068 \\
GPT-4o & 0.762 & 0.222 & 0.009 & 0.007 \\
GPT-5 Nano & 0.761 & 0.163 & 0.068 & 0.008 \\
GPT-5.2 Pro & 0.756 & 0.221 & 0.013 & 0.010 \\
Claude 3 Haiku & 0.759 & 0.120 & 0.116 & 0.004 \\
Claude 3.7 Sonnet & 0.731 & 0.232 & 0.007 & 0.030 \\
Claude Opus 4.5 & 0.708 & 0.225 & 0.006 & 0.061 \\
\bottomrule
\end{tabular}
\end{table}

\subsection{Results}
\label{sec:results}

\paragraph{LLMs reach near-ceiling F1.}
\Cref{tab:main-results} reports the main results. The strongest LLMs reach near-ceiling performance (98--99\% F1), substantially outperforming the rule-based baseline (91.33\% F1) used in production. GPT-4o achieves the highest score (98.95\% F1), and the best open-source model (DeepSeek-R1-Distill-Qwen-14B, 98.23\% F1) closely matches frontier proprietary models such as GPT-5.2 Pro (98.75\% F1).

\paragraph{Prompt optimization provides minor gains.}
MIPROv2 prompt optimization yields modest, model-dependent improvements: Llama-3.1-8B improves from 94.05\% to 95.94\% F1 (+1.9), while frontier proprietary models change negligibly. The default and optimized prompts are shown side-by-side in \Cref{fig:prompt-comparison} (Appendix~\ref{app:prompt-comparison}).

\paragraph{Few-shot examples do not help once the prompt is optimized.}
Llama-3.1-8B with 2-, 4-, and 8-shot demonstrations matches or underperforms its optimized zero-shot baseline (95.46\%, 95.64\%, 95.94\% vs.\ 95.94\% F1); GPT-5 Nano shows the same pattern. In-context demonstrations appear to introduce inductive biases that compete with the optimized instruction.

\paragraph{Model size vs.\ throughput.}
DeepSeek-14B slightly outperforms Llama-3.1-8B (98.23\% vs.\ 95.94\% F1) but takes roughly $3\times$ longer per example ($\sim$70s vs.\ $\sim$25s); Llama-3.1-8B is the better accuracy-throughput point for at-scale deployment.

\begin{figure*}[!t]
\centering
\includegraphics[width=1.0\textwidth]{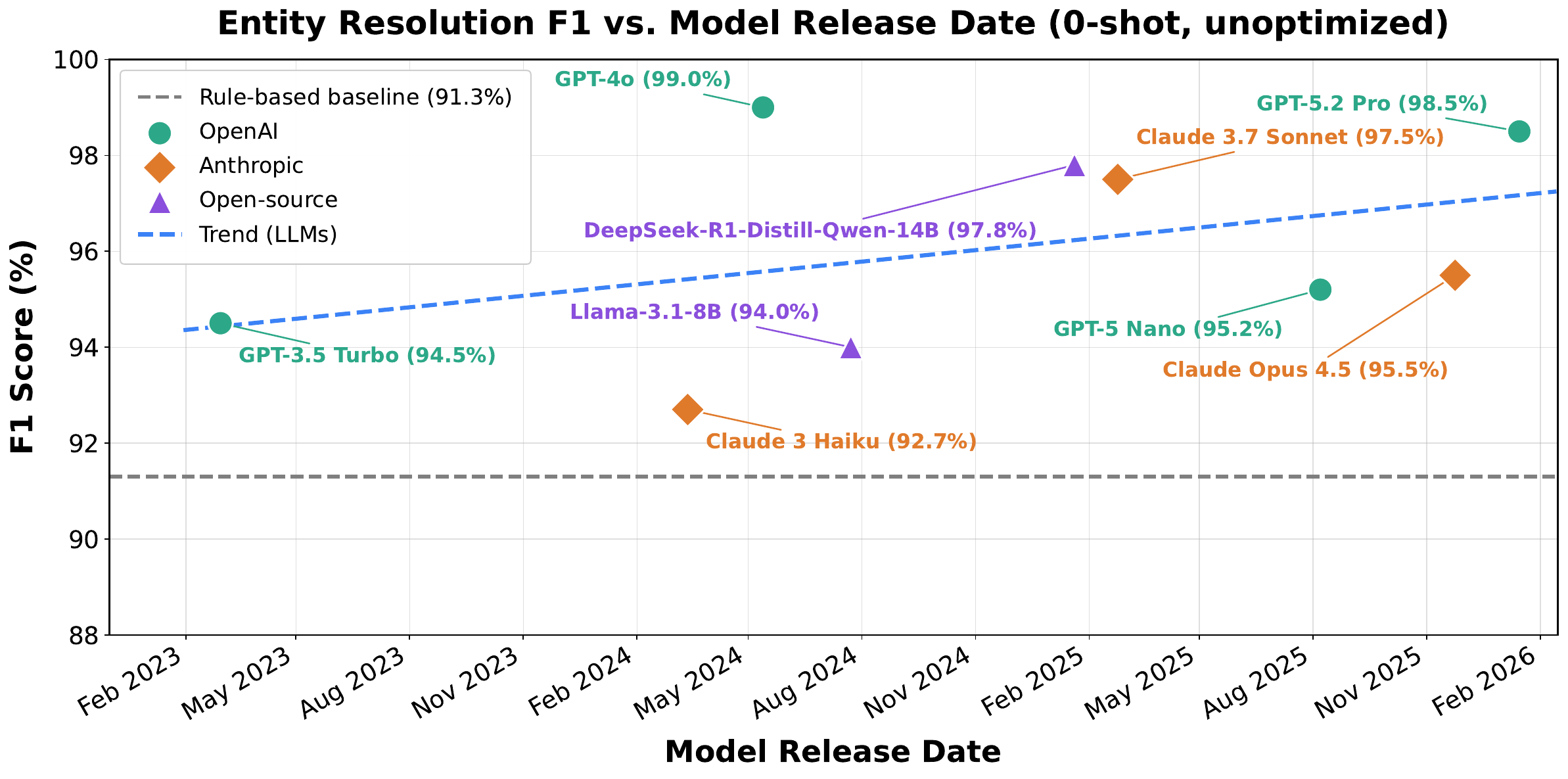}
\caption{Unoptimized (0-shot) F1 versus model release date. The dashed gray line is the Nomenklatura rule baseline (91.3\% F1); the blue dashed line is a linear fit across LLMs. The LLM-rule gap widened from $\sim$2 F1 points in 2023 to over 7 by late 2025.}
\label{fig:f1-vs-release}
\end{figure*}

\paragraph{Performance vs.\ release date.}
\Cref{fig:f1-vs-release} shows 0-shot F1 rising from 92--94\% (early 2023) to 98--99\% (late 2025), suggesting that pairwise matching is approaching a practical ceiling and that further gains are likelier to come from blocking, clustering, and uncertainty-aware review than from pairwise prompt tuning.

\subsection{Failure modes}
\label{sec:failure-modes}

The per-method confusion matrices in \Cref{tab:confusion} reveal two distinct error profiles. RegressionV1 has near-perfect recall but many false positives, reflecting OpenSanctions' operational preference for catching every potential match at the cost of analyst review (threshold 0.15). LLM errors instead concentrate in two narrow modes: cross-script transliteration differences (Latin / Cyrillic / Arabic) that obscure genuine matches, and over-weighting of minor data-entry noise (off-by-one dates or identifiers) into spurious negatives under the conflict-focused prompt.

\section{Limitations}
\label{sec:limitations}

We note three limitations. (1) Labels reflect analyst judgments under incomplete evidence, so model-label disagreements may capture labeling uncertainty rather than prediction errors. (2) Evaluation sample sizes vary across models due to compute cost; all are label-stratified, but smaller ones carry greater statistical uncertainty. (3) Most OpenSanctions sources are public, so overlap with frontier-model training corpora cannot be fully excluded; the released \texttt{first\_seen} timestamps support a temporal hold-out for future leakage probes.

\section{Conclusion}

We released OpenSanctions Pairs, the first large-scale public benchmark for entity matching on sanctions OSINT data: 756K analyst-labeled pairs over 1.0M entities aggregated from 293 sources across 45 jurisdictions, multi-script and multi-schema, with set-valued partial-evidence attributes that mirror operational compliance workflows. Reference baselines establish that the benchmark is non-trivial yet within reach of current LLMs (91.3\% rule baseline vs.\ 98--99\% LLM ceiling), with complementary failure modes (rule over-matching vs.\ LLM cross-script transliteration) that point to future work on blocking, clustering, and uncertainty-aware review.

Sanctions matching has until now been evaluated almost entirely inside vendor labs, leaving research progress invisible to the field. OpenSanctions Pairs gives the community a shared target on the same data that banks, regulators, NGOs, and investigative journalists rely on to flag sanctioned individuals and shell companies. Each point of recall closes a route by which designated parties evade screening through aliases or transliterations; each point of precision spares analysts and legitimate counterparties from thousands of false hits.

% \section*{Acknowledgements}

\section*{Impact statement}

This work releases a public benchmark for entity matching on real-world sanctions data, supporting realistic evaluation of LLMs on multi-script, cross-jurisdictional compliance tasks. Surfacing the complementary failure modes of rule-based and learned matchers encourages safer human-in-the-loop deployment and reduces unnecessary false positives. We acknowledge dual-use risk: matching infrastructure that aids compliance can also support adversarial profiling, partly mitigated by the CC-BY-NC license and analyst-judgment caveats accompanying the release.

\clearpage

\bibliographystyle{abbrvnat}
\bibliography{main}

\begin{thebibliography}{43}
\providecommand{\natexlab}[1]{#1}
\providecommand{\url}[1]{\texttt{#1}}
\expandafter\ifx\csname urlstyle\endcsname\relax
  \providecommand{\doi}[1]{doi: #1}\else
  \providecommand{\doi}{doi: \begingroup \urlstyle{rm}\Url}\fi

\bibitem[Barlaug and Gulla(2021)]{Barlaug_2021}
N.~Barlaug and J.~A. Gulla.
\newblock Neural networks for entity matching: A survey.
\newblock \emph{ACM Trans. Knowl. Discov. Data}, 15\penalty0 (3), Apr. 2021.
\newblock ISSN 1556-4681.
\newblock \doi{10.1145/3442200}.
\newblock URL \url{https://doi.org/10.1145/3442200}.

\bibitem[Christen(2012)]{Christen_2012}
P.~Christen.
\newblock \emph{Data Matching: Concepts and Techniques for Record Linkage,
  Entity Resolution, and Duplicate Detection}.
\newblock Springer Berlin Heidelberg, 2012.
\newblock ISBN 9783642311642.
\newblock \doi{10.1007/978-3-642-31164-2}.
\newblock URL \url{http://dx.doi.org/10.1007/978-3-642-31164-2}.

\bibitem[Christophides et~al.(2020)Christophides, Efthymiou, Palpanas,
  Papadakis, and Stefanidis]{Christophides_2020}
V.~Christophides, V.~Efthymiou, T.~Palpanas, G.~Papadakis, and K.~Stefanidis.
\newblock End-to-end entity resolution for big data: A survey, 2020.
\newblock URL \url{https://arxiv.org/abs/1905.06397}.

\bibitem[{DSPy}(n.d.)]{dspy_MIPROv2}
{DSPy}.
\newblock Miprov2 --- dspy api optimizer.
\newblock \url{https://dspy.ai/api/optimizers/MIPROv2/}, n.d.
\newblock URL \url{https://dspy.ai/api/optimizers/MIPROv2/}.
\newblock Accessed: 2026-01-29.

\bibitem[Elmagarmid et~al.(2007)Elmagarmid, Ipeirotis, and
  Verykios]{Elmagarmid_2007}
A.~K. Elmagarmid, P.~G. Ipeirotis, and V.~S. Verykios.
\newblock Duplicate record detection: A survey.
\newblock \emph{IEEE Trans. on Knowl. and Data Eng.}, 19\penalty0 (1):\penalty0
  1–16, Jan. 2007.
\newblock ISSN 1041-4347.

\bibitem[Fan et~al.(2023)Fan, Han, Fan, Chai, Tang, Li, and Du]{Fan_2023}
M.~Fan, X.~Han, J.~Fan, C.~Chai, N.~Tang, G.~Li, and X.~Du.
\newblock Cost-effective in-context learning for entity resolution: A design
  space exploration, 2023.
\newblock URL \url{https://arxiv.org/abs/2312.03987}.

\bibitem[Felbermayr et~al.(2020)Felbermayr, Kirilakha, Syropoulos, Yalcin, and
  Yotov]{Felbermayr_2020}
G.~Felbermayr, A.~Kirilakha, C.~Syropoulos, E.~Yalcin, and Y.~V. Yotov.
\newblock The global sanctions data base.
\newblock \emph{European Economic Review}, 129:\penalty0 103561, 2020.
\newblock ISSN 0014-2921.
\newblock \doi{https://doi.org/10.1016/j.euroecorev.2020.103561}.
\newblock URL
  \url{https://www.sciencedirect.com/science/article/pii/S0014292120301914}.

\bibitem[Fellegi and Sunter(1969)]{Fellegi_Sunter_1969}
I.~P. Fellegi and A.~B. Sunter.
\newblock A theory for record linkage.
\newblock \emph{Journal of the American Statistical Association}, 64\penalty0
  (328):\penalty0 1183--1210, 1969.
\newblock \doi{10.1080/01621459.1969.10501049}.
\newblock URL
  \url{https://www.tandfonline.com/doi/abs/10.1080/01621459.1969.10501049}.

\bibitem[Fu et~al.(2026)Fu, Tang, Khan, Mehrotra, Ke, and Gao]{Fu_2026}
J.~Fu, H.~Tang, A.~Khan, S.~Mehrotra, X.~Ke, and Y.~Gao.
\newblock In-context clustering-based entity resolution with large language
  models: A design space exploration.
\newblock In \emph{Proceedings of the ACM on Management of Data (SIGMOD)},
  2026.
\newblock URL \url{https://arxiv.org/abs/2506.02509}.

\bibitem[Getoor and Machanavajjhala(2012)]{Getoor_2012}
L.~Getoor and A.~Machanavajjhala.
\newblock Entity resolution: Theory, practice \& open challenges.
\newblock \emph{Proceedings of the VLDB Endowment}, 5:\penalty0 2018--2019, 08
  2012.
\newblock \doi{10.14778/2367502.2367564}.

\bibitem[Hernández and Stolfo(1995)]{Hernandez_1995}
M.~A. Hernández and S.~J. Stolfo.
\newblock The merge/purge problem for large databases.
\newblock In \emph{Proceedings of the 1995 ACM SIGMOD international conference
  on Management of data - SIGMOD ’95}, SIGMOD ’95, page 127–138. ACM
  Press, 1995.
\newblock \doi{10.1145/223784.223807}.
\newblock URL \url{http://dx.doi.org/10.1145/223784.223807}.

\bibitem[Kim and Yang(2024)]{Kim_2024}
S.~Kim and S.~Yang.
\newblock Accuracy improvement in financial sanction screening: is natural
  language processing the solution?
\newblock \emph{Frontiers in Artificial Intelligence}, 7, Nov. 2024.
\newblock ISSN 2624-8212.
\newblock \doi{10.3389/frai.2024.1374323}.
\newblock URL \url{http://dx.doi.org/10.3389/frai.2024.1374323}.

\bibitem[Konda et~al.(2016)Konda, Das, Suganthan G.~C., Doan, Ardalan, Ballard,
  Li, Panahi, Zhang, Naughton, Prasad, Krishnan, Deep, and
  Raghavendra]{Konda_2016}
P.~Konda, S.~Das, P.~Suganthan G.~C., A.~Doan, A.~Ardalan, J.~R. Ballard,
  H.~Li, F.~Panahi, H.~Zhang, J.~Naughton, S.~Prasad, G.~Krishnan, R.~Deep, and
  V.~Raghavendra.
\newblock Magellan: toward building entity matching management systems.
\newblock \emph{Proc. VLDB Endow.}, 9\penalty0 (12):\penalty0 1197–1208, Aug.
  2016.
\newblock ISSN 2150-8097.
\newblock \doi{10.14778/2994509.2994535}.
\newblock URL \url{https://doi.org/10.14778/2994509.2994535}.

\bibitem[K\"{o}pcke et~al.(2010{\natexlab{a}})K\"{o}pcke, Thor, and
  Rahm]{Kopcke_2010a}
H.~K\"{o}pcke, A.~Thor, and E.~Rahm.
\newblock Evaluation of entity resolution approaches on real-world match
  problems.
\newblock \emph{Proceedings of the VLDB Endowment}, 3\penalty0
  (1–2):\penalty0 484–493, Sept. 2010{\natexlab{a}}.
\newblock ISSN 2150-8097.
\newblock \doi{10.14778/1920841.1920904}.
\newblock URL \url{http://dx.doi.org/10.14778/1920841.1920904}.

\bibitem[K\"{o}pcke et~al.(2010{\natexlab{b}})K\"{o}pcke, Thor, and
  Rahm]{Kopcke_2010b}
H.~K\"{o}pcke, A.~Thor, and E.~Rahm.
\newblock Learning-based approaches for matching web data entities.
\newblock \emph{IEEE Internet Computing}, 14\penalty0 (4):\penalty0 23–31,
  July 2010{\natexlab{b}}.
\newblock ISSN 1089-7801.
\newblock \doi{10.1109/mic.2010.58}.
\newblock URL \url{http://dx.doi.org/10.1109/MIC.2010.58}.

\bibitem[Li et~al.(2024{\natexlab{a}})Li, Feng, Li, Hao, Zhang, and
  Song]{Li_2024_arxiv}
H.~Li, L.~Feng, S.~Li, F.~Hao, C.~J. Zhang, and Y.~Song.
\newblock On leveraging large language models for enhancing entity resolution:
  A cost-efficient approach, 2024{\natexlab{a}}.
\newblock URL \url{https://arxiv.org/abs/2401.03426}.

\bibitem[Li et~al.(2024{\natexlab{b}})Li, Li, Hao, Zhang, Song, and
  Chen]{Li_2024_acm}
H.~Li, S.~Li, F.~Hao, C.~J. Zhang, Y.~Song, and L.~Chen.
\newblock Booster: Leveraging large language models for enhancing entity
  resolution.
\newblock In \emph{Companion Proceedings of the ACM Web Conference 2024}, WWW
  '24, page 1043–1046, New York, NY, USA, 2024{\natexlab{b}}. Association for
  Computing Machinery.
\newblock ISBN 9798400701726.
\newblock \doi{10.1145/3589335.3651245}.
\newblock URL \url{https://doi.org/10.1145/3589335.3651245}.

\bibitem[Li et~al.(2020)Li, Li, Suhara, Doan, and Tan]{Li_2020}
Y.~Li, J.~Li, Y.~Suhara, A.~Doan, and W.-C. Tan.
\newblock Deep entity matching with pre-trained language models.
\newblock \emph{Proceedings of the VLDB Endowment}, 14\penalty0 (1):\penalty0
  50–60, Sept. 2020.
\newblock ISSN 2150-8097.
\newblock \doi{10.14778/3421424.3421431}.
\newblock URL \url{http://dx.doi.org/10.14778/3421424.3421431}.

\bibitem[Montenarh and Marsden(2024)]{Montenarh_2024}
J.~Montenarh and S.~Marsden.
\newblock Unmasking the oligarchs – using open source data to detect
  sanctions violations.
\newblock \emph{Journal of Economic Criminology}, 3:\penalty0 100055, Mar.
  2024.
\newblock ISSN 2949-7914.
\newblock \doi{10.1016/j.jeconc.2024.100055}.
\newblock URL \url{http://dx.doi.org/10.1016/j.jeconc.2024.100055}.

\bibitem[Moslemi et~al.(2026)Moslemi, Mousavi, Behkamal, and
  Milani]{Moslemi_2026}
M.~H. Moslemi, A.~Mousavi, B.~Behkamal, and M.~Milani.
\newblock Heterogeneity in entity matching: A survey and experimental analysis,
  2026.
\newblock URL \url{https://arxiv.org/abs/2508.08076}.
\newblock Revised February 2026.

\bibitem[Mudgal et~al.(2018{\natexlab{a}})Mudgal, Li, Rekatsinas, Doan, Park,
  Krishnan, Deep, Arcaute, and Raghavendra]{Mudgal_2018}
S.~Mudgal, H.~Li, T.~Rekatsinas, A.~Doan, Y.~Park, G.~Krishnan, R.~Deep,
  E.~Arcaute, and V.~Raghavendra.
\newblock Deep learning for entity matching: A design space exploration.
\newblock In \emph{Proceedings of the 2018 International Conference on
  Management of Data}, SIGMOD '18, page 19–34, New York, NY, USA,
  2018{\natexlab{a}}. Association for Computing Machinery.
\newblock ISBN 9781450347037.
\newblock \doi{10.1145/3183713.3196926}.
\newblock URL \url{https://doi.org/10.1145/3183713.3196926}.

\bibitem[Mudgal et~al.(2018{\natexlab{b}})Mudgal, Li, Rekatsinas, Doan, Park,
  Krishnan, Deep, Arcaute, and Raghavendra]{mudgal_deepmatcher_datasets}
S.~Mudgal, H.~Li, T.~Rekatsinas, A.~Doan, Y.~Park, G.~Krishnan, R.~Deep,
  E.~Arcaute, and V.~Raghavendra.
\newblock Datasets for the deepmatcher paper.
\newblock
  \url{https://github.com/anhaidgroup/deepmatcher/blob/master/Datasets.md},
  2018{\natexlab{b}}.
\newblock Used in SIGMOD 2018 paper ``Deep Learning for Entity Matching'';
  accessed 2026-01-28.

\bibitem[Narayan et~al.(2022)Narayan, Chami, Orr, Arora, and Ré]{Narayan_2022}
A.~Narayan, I.~Chami, L.~Orr, S.~Arora, and C.~Ré.
\newblock Can foundation models wrangle your data?, 2022.
\newblock URL \url{https://arxiv.org/abs/2205.09911}.

\bibitem[{OpenSanctions}(2021)]{opensanctions_dedup}
{OpenSanctions}.
\newblock How we deduplicate companies and people across data sources, 2021.
\newblock URL
  \url{https://www.opensanctions.org/articles/2021-11-11-deduplication/}.
\newblock Accessed: 2025-01-15.

\bibitem[OpenSanctions(2025)]{nomenklatura}
OpenSanctions.
\newblock {Nomenklatura}: Framework and command-line tools for integrating
  followthemoney data streams from multiple sources.
\newblock \url{https://github.com/opensanctions/nomenklatura/}, 2025.
\newblock Accessed: 2026-01-20.

\bibitem[{OpenSanctions}(2025)]{opensanctions_pairs}
{OpenSanctions}.
\newblock Matcher training data, 2025.
\newblock URL \url{https://www.opensanctions.org/docs/opensource/pairs/}.
\newblock Accessed: 2025-01-15.

\bibitem[Ornstein(2025)]{Ornstein_2025}
J.~T. Ornstein.
\newblock Probabilistic record linkage using pretrained text embeddings.
\newblock \emph{Political Analysis}, 34\penalty0 (2):\penalty0 205–216, 2025.
\newblock \doi{10.1017/pan.2025.10016}.

\bibitem[O’Hare et~al.(2019)O’Hare, Jurek-Loughrey, and Campos]{OHare_2018}
K.~O’Hare, A.~Jurek-Loughrey, and C.~d. Campos.
\newblock \emph{A Review of Unsupervised and Semi-supervised Blocking Methods
  for Record Linkage}, page 79–105.
\newblock Springer International Publishing, 2019.
\newblock ISBN 9783030018726.
\newblock \doi{10.1007/978-3-030-01872-6_4}.
\newblock URL \url{http://dx.doi.org/10.1007/978-3-030-01872-6_4}.

\bibitem[Papadakis et~al.(2020)Papadakis, Skoutas, Thanos, and
  Palpanas]{Papadakis_2020}
G.~Papadakis, D.~Skoutas, E.~Thanos, and T.~Palpanas.
\newblock A survey of blocking and filtering techniques for entity resolution,
  2020.
\newblock URL \url{https://arxiv.org/abs/1905.06167}.

\bibitem[Peeters(2025)]{Peeters_2025}
R.~Peeters.
\newblock \emph{Entity Matching using Deep Neural Networks: From Discriminative
  Pre-trained Language Models to Generative Large Language Models}.
\newblock PhD thesis, University of Mannheim, 2025.
\newblock URL \url{https://madoc.bib.uni-mannheim.de/69425/}.

\bibitem[Peeters and Bizer(2023)]{Peeters_2023}
R.~Peeters and C.~Bizer.
\newblock Using chatgpt for entity matching, 2023.
\newblock URL \url{https://arxiv.org/abs/2305.03423}.

\bibitem[Peeters et~al.(2023)Peeters, Der, and Bizer]{Peeters_2023_WDC}
R.~Peeters, R.~C. Der, and C.~Bizer.
\newblock Wdc products: A multi-dimensional entity matching benchmark, 2023.
\newblock URL \url{https://arxiv.org/abs/2301.09521}.

\bibitem[Peeters et~al.(2024)Peeters, Steiner, and Bizer]{Peeters_2024}
R.~Peeters, A.~Steiner, and C.~Bizer.
\newblock Entity matching using large language models, 2024.
\newblock URL \url{https://arxiv.org/abs/2310.11244}.

\bibitem[Primpeli et~al.(2019)Primpeli, Peeters, and
  Bizer]{Primpeli_2019_WDC_LSPC}
A.~Primpeli, R.~Peeters, and C.~Bizer.
\newblock The {WDC} training dataset and gold standard for large-scale product
  matching.
\newblock In \emph{Companion of the 2019 World Wide Web Conference (WWW '19)},
  pages 381--386. ACM, 2019.
\newblock \doi{10.1145/3308560.3316609}.
\newblock URL \url{https://webdatacommons.org/largescaleproductcorpus/v2/}.

\bibitem[Rahm and Leipzig(2019)]{Rahm_entity_resolution_datasets_2019}
E.~Rahm and D.~G. Leipzig.
\newblock Benchmark datasets for entity resolution, 2019.
\newblock URL
  \url{https://dbs.uni-leipzig.de/research/projects/benchmark-datasets-for-entity-resolution}.
\newblock Dataset collection released 2010--2019; accessed 2026-01-28.

\bibitem[Sadinle(2016)]{Sadinle_2016}
M.~Sadinle.
\newblock Bayesian estimation of bipartite matchings for record linkage, 2016.
\newblock URL \url{https://arxiv.org/abs/1601.06630}.

\bibitem[Steiner et~al.(2025)Steiner, Peeters, and Bizer]{Steiner_2025}
A.~Steiner, R.~Peeters, and C.~Bizer.
\newblock Fine-tuning large language models for entity matching, 2025.
\newblock URL \url{https://arxiv.org/abs/2409.08185}.

\bibitem[Steorts et~al.(2015)Steorts, Hall, and Fienberg]{Steorts_2015}
R.~C. Steorts, R.~Hall, and S.~E. Fienberg.
\newblock A bayesian approach to graphical record linkage and de-duplication,
  2015.
\newblock URL \url{https://arxiv.org/abs/1312.4645}.

\bibitem[Sterling and Meijer(2018)]{Sterling_2018}
T.~Sterling and B.~H. Meijer.
\newblock Dutch bank ing fined \$900 million for failing to spot money
  laundering.
\newblock
  \url{https://www.reuters.com/article/us-ing-groep-settlement-money-laundering-idUSKCN1LK0PE/},
  September 4 2018.
\newblock Accessed: 2026-01-13.

\bibitem[Tang et~al.(2022)Tang, Zuo, Cao, and Madden]{Tang_2022}
J.~Tang, Y.~Zuo, L.~Cao, and S.~Madden.
\newblock Generic entity resolution models.
\newblock In \emph{NeurIPS 2022 First Table Representation Workshop}, 2022.
\newblock URL \url{https://openreview.net/forum?id=tRkVo1jMas}.

\bibitem[Wang et~al.(2025)Wang, Chen, Lin, Chen, Han, Sun, Wang, and
  Zeng]{Wang_2025}
T.~Wang, X.~Chen, H.~Lin, X.~Chen, X.~Han, L.~Sun, H.~Wang, and Z.~Zeng.
\newblock Match, compare, or select? an investigation of large language models
  for entity matching.
\newblock In O.~Rambow, L.~Wanner, M.~Apidianaki, H.~Al-Khalifa, B.~D. Eugenio,
  and S.~Schockaert, editors, \emph{Proceedings of the 31st International
  Conference on Computational Linguistics}, pages 96--109, Abu Dhabi, UAE, Jan.
  2025. Association for Computational Linguistics.
\newblock URL \url{https://aclanthology.org/2025.coling-main.8/}.

\bibitem[Zeakis et~al.(2026)Zeakis, Papadakis, Skoutas, and
  Koubarakis]{Zeakis_2026}
A.~Zeakis, G.~Papadakis, D.~Skoutas, and M.~Koubarakis.
\newblock Distiller: Knowledge distillation in entity resolution with large
  language models, 2026.
\newblock URL \url{https://arxiv.org/abs/2602.05452}.

\bibitem[Zhang et~al.(2025)Zhang, Groth, Calixto, and Schelter]{Zhang_2025}
Z.~Zhang, P.~Groth, I.~Calixto, and S.~Schelter.
\newblock {ANYMATCH} {\textendash} efficient zero-shot entity matching with a
  small language model.
\newblock In \emph{AAAI 2025 Workshop on Preparing Good Data for Generative AI:
  Challenges and Approaches}, 2025.
\newblock URL \url{https://openreview.net/forum?id=Nees1OD5td}.

\end{thebibliography}

\clearpage

%%%%%%%%%%%%%%%%%%%%%%%%%%%%%%%%%%%%%%%%%%%%%%%%%%%%%%%%%%%%
%%%%%%%%%%%%%%%%%%%%%%%%%%%%%%%%%%%%%%%%%%%%%%%%%%%%%%%%%%%%
% APPENDIX
%%%%%%%%%%%%%%%%%%%%%%%%%%%%%%%%%%%%%%%%%%%%%%%%%%%%%%%%%%%%
%%%%%%%%%%%%%%%%%%%%%%%%%%%%%%%%%%%%%%%%%%%%%%%%%%%%%%%%%%%%

\appendix

\section{Prompts}
\label{app:prompts}
This appendix contains the prompts used for LLM-based entity resolution experiments: the manually designed default prompt, the DSPy MIPROv2-optimized prompt, and the structured output schema and entity formatting used at inference.

\subsection{Default prompt}
\label{app:default-prompt}
Our baseline prompt was manually designed around a conflict-detection framing, achieving 94.95\% F1 on GPT-5 Nano. The key insight is framing entity resolution as contradiction detection rather than similarity matching.

\begin{lstlisting}[basicstyle=\small\ttfamily,breaklines=true,frame=single,title={\textbf{System Prompt.}}]
You are an expert entity resolution system. Your task is to determine if two entity records refer to the same real-world entity.

Primary task: Identify CONFLICTS, not similarities.
- Name variations (transliterations, nicknames, titles) are common
- Missing fields are normal - absence of data is NOT evidence of difference
- Same entity often appears across multiple sources with variations

Decision Process:
1. Look for CONTRADICTORY evidence (different dates, conflicting IDs, incompatible attributes)
2. If NO contradictions found -> POSITIVE (same entity)
3. Only NEGATIVE if explicit conflicts exist

The DEFAULT is POSITIVE unless you find proof of difference.
\end{lstlisting}

\begin{lstlisting}[basicstyle=\small\ttfamily,breaklines=true,frame=single,title={\textbf{User Prompt Template.}}]
Determine if these two entity records refer to the same real-world entity.

=== Entity A ===
{entity_a}

=== Entity B ===
{entity_b}

Classify as:
- POSITIVE: Same entity (no conflicts found)
- NEGATIVE: Different entities (explicit conflicts exist)

Respond with a JSON object containing:
- "classification": "positive" or "negative"
- "reasoning": Brief explanation focusing on conflicts or lack thereof
\end{lstlisting}

\subsection{Optimized prompt (DeepSeek-R1-Distill-Qwen-14B)}
\label{app:optimized-prompt}
This prompt was automatically generated by DSPy MIPROv2 optimization, achieving 98.23\% F1 with DeepSeek-R1-Distill-Qwen-14B. The optimizer discovered a more concise framing that retains the conflict-detection principle.

\begin{lstlisting}[basicstyle=\small\ttfamily,breaklines=true,frame=single,title={\textbf{Instructions.}}]
Given two entity records, determine if they refer to the same
real-world entity. Analyze for contradictions in attributes like
dates, IDs, or other details. Consider common variations in names
and missing data. Classify as 'positive' if no conflicts are found,
otherwise 'negative'. Provide a clear reasoning for your classification.
\end{lstlisting}

\subsection{Prompt comparison: manual vs.\ MIPROv2}
\label{app:prompt-comparison}

\begin{figure*}[ht]
\centering
\begin{minipage}[c]{0.49\textwidth}
\begin{tcolorbox}[
  colback=gray!5,
  colframe=gray!50,
  boxrule=0.5pt,
  arc=2pt,
  title={\small\textbf{Before Optimization (Manual Design)}}
]
\scriptsize\ttfamily
\textbf{System:} You are an expert entity resolution system. Your task is to determine if two entity records refer to the same real-world entity.\\[0.3em]
Primary task: Identify CONFLICTS, not similarities.\\
- Name variations (transliterations, nicknames, titles) are common\\
- Missing fields are normal - absence of data is NOT evidence of difference\\
- Same entity often appears across multiple sources with variations\\[0.3em]
Decision Process:\\
1. Look for CONTRADICTORY evidence (different dates, conflicting IDs, incompatible attributes)\\
2. If NO contradictions found -> POSITIVE (same entity)\\
3. Only NEGATIVE if explicit conflicts exist\\[0.3em]
The DEFAULT is POSITIVE unless you find proof of difference.\\[0.5em]
\textbf{User:} Determine if these two entity records refer to the same real-world entity.\\[0.3em]
=== Entity A ===\\
\{entity\_a\}\\[0.3em]
=== Entity B ===\\
\{entity\_b\}\\[0.3em]
Classify as POSITIVE (same entity) or NEGATIVE (different entities).\\
Respond with JSON: \{"classification": ..., "reasoning": ...\}
\end{tcolorbox}
\end{minipage}
\hfill
\begin{minipage}[c]{0.49\textwidth}
\begin{tcolorbox}[
  colback=gray!5,
  colframe=gray!50,
  boxrule=0.5pt,
  arc=2pt,
  title={\small\textbf{After Optimization (MIPROv2)}}
]
\scriptsize\ttfamily
\textbf{Instructions:} Given two entity records, determine if they refer to the same real-world entity. Analyze for contradictions in attributes like dates, IDs, or other details. Consider common variations in names and missing data. Classify as 'positive' if no conflicts are found, otherwise 'negative'. Provide a clear reasoning for your classification.\\[0.5em]
\textbf{Input:}\\
Entity A: \{entity\_a\}\\
Entity B: \{entity\_b\}\\[0.3em]
\textbf{Output:}\\
Reasoning: \{reasoning\}\\
Classification: \{positive|negative\}
\end{tcolorbox}
\end{minipage}
\caption{Manual prompt (left, 170 words) vs.\ the MIPROv2-optimized prompt (right, 50 words). The optimizer discovered a more concise framing while preserving the conflict-detection principle.}
\label{fig:prompt-comparison}
\end{figure*}

\subsection{Structured output schema}
\label{app:output-schema}
We use JSON structured output to ensure consistent parsing across all LLM experiments:

\begin{lstlisting}[basicstyle=\small\ttfamily,breaklines=true,frame=single]
{
  "type": "object",
  "properties": {
    "classification": {
      "type": "string",
      "enum": ["positive", "negative"]
    },
    "reasoning": "string"
  },
  "required": ["classification", "reasoning"]
}
\end{lstlisting}

%%%%%%%%%%%%%%%%%%%%%%%%%%%%%%%%%%%%%%%%%%%%%%%%%%%%%%%%%%%%

% NeurIPS submission checklist omitted from the arXiv preprint.
% \newpage
% \input{checklist.tex}

\end{document}